\documentclass[aps,prb,twocolumn,superscriptaddress]{revtex4-2}

\usepackage{amsmath,amssymb}
\usepackage{bm}
\usepackage{physics}
\usepackage{hyperref}
\usepackage{graphicx}
\usepackage{xcolor}

\begin{document}

\title{Chiral vortical effect and boundary-induced vortical pumping in finite Weyl systems}

\author{Boqun Song}
\affiliation{Department of Physics, University of Houston, Houston, Texas 77204, USA}
\affiliation{Texas Center for Superconductivity, University of Houston, Houston, Texas 77204, USA}

\author{Pavan Hosur}
\affiliation{Department of Physics, University of Houston, Houston, Texas 77204, USA}
\affiliation{Texas Center for Superconductivity, University of Houston, Houston, Texas 77204, USA}
\affiliation{Department of Electrical and Computer Engineering, Rice University, Houston, Texas, 77005, USA}

\begin{abstract}
The chiral vortical effect (CVE) -- an axial current driven by rotation in chiral matter -- appears in systems ranging from relativistic fluids to Weyl semimetals. We present a quantum treatment of a rotating Weyl fermion in a finite cylinder that reproduces the on-axis current density obtained by Vilenkin in an infinite system, while revealing two new phenomena. First, we find that the cross-sectional averaged current density vanishes in the thermodynamic limit, establishing that the bulk vortical response is purely a magnetization current. Second, for spin-polarized boundary conditions, we uncover a robust family of chiral modes that transport axial charge, $\Delta Q=\chi N^2\Delta\theta/4\pi$, under rotation by angle $\Delta\theta$, where $\chi$ is the Weyl node chirality and $N$ is the number of chiral modes. The pump is independent of temperature, Fermi level, and Weyl velocities, but depends on the UV-sensitive number $N$. These results clarify the quantum structure of the bulk CVE and reveal a boundary-enforced chiral spectral structure underlying vortical response in Weyl systems.
\end{abstract}

\maketitle

\section{Introduction}

The chiral vortical effect (CVE) \cite{Vilenkin1979,Vilenkin1980,Vilenkin1978a,VILENKIN1978b,LandsteinerAnomaly,Nanda2023,KirilinCVESF,KhaidukovCVEFL,Loganayagam2012,SadofyevChiralHydroNotes,Chen2024chiralkinetic} is an axial current generated by vorticity or rotation in chiral matter. It appears in systems ranging from parity-violating heavy-ion collisions and neutrino transport near rotating black holes \cite{Vilenkin1979} to electron transport in Weyl semimetals \cite{LandsteinerAnomaly,Nanda2023,Chen2024chiralkinetic}. An exact quantum treatment of rotating chiral fermions was given by Vilenkin \cite{Vilenkin1979,Vilenkin1980}, who derived the local axial current density on the rotation axis. More recently, the CVE has been studied extensively using semiclassical kinetic theory, hydrodynamics, and linear response \cite{Chen2014,StephanovKineticTheory,Loganayagam2012,LandsteinerAnomaly,KhaidukovCVEFL,KirilinCVESF,Shitade2020,Stone2018MixedExpansion,Nanda2023,Chen2024chiralkinetic}. These complementary approaches have established the bulk vortical response in a variety of physical settings and clarified its connection to anomalies and transport.

At the same time, nearly all existing treatments concern infinite bulk systems. In contrast, finite Weyl systems possess boundaries that can host electronic states absent in the bulk, most notably Fermi arcs. Such states naturally raise the question of whether boundaries merely modify the known bulk CVE or give rise to qualitatively new forms of vortical response. More generally, a finite geometry provides access to aspects of the vortical response that are absent from bulk descriptions, including the distinction between local and transport currents and the role of boundary states.

In this work we answer this question by developing an exact quantum treatment of a rotating Weyl fermion in a finite system. Starting from the Weyl equation in a rotating frame, we assume a non-equilibrium steady-state distribution that appears thermal to a co-rotating observer. We recover the established current density locally on the rotation axis obtained by Vilenkin \cite{Vilenkin1979}. In addition, we find that the transport current density carried by the bulk states vanishes in the thermodynamic limit, implying that the \emph{bulk} CVE current density is of purely magnetization origin.

Beyond the bulk response, the presence of a boundary places the system in a qualitatively distinct regime, where boundary-supported modes can contribute to vortical response. While general boundary conditions render the surface state wavefunctions intractable, a fully spin-polarized boundary admits an exactly solvable family of critical states with remarkable properties: (i) they are 1D chiral modes; (ii) their wavefunctions have simple, holomorphic closed forms for arbitrary boundary shapes and sizes; and (iii) under rotation, they pump axial charge:
\begin{equation}
\label{eq:charge-pump}
\Delta Q=\chi\frac{\Delta\theta}{4\pi}N^2
\end{equation}
where $\Delta Q$ is the charge transported during a rotation by angle $\Delta\theta$, $\chi$ is the handedness of the Weyl node, and $N\in\mathbb{Z}\geq1$ is the number of chiral modes, determined by a UV cutoff on angular momentum. In particular, a $4\pi$ global rotation pumps an integer number ($N^2$) of particles. We stress that the pump is non-universal as $N$ depends on microscopic model details. Nonetheless, this boundary-induced pumping lies beyond existing semiclassical and bulk quantum treatments \cite{SonSpivakWeylAnomaly,StephanovKineticTheory,Chen2014,Nanda2023,Chen2024chiralkinetic,Loganayagam2012,LandsteinerAnomaly,KhaidukovCVEFL,KirilinCVESF,Son2012,Shitade2020,Kharzeev:2016vn,Stone2018MixedExpansion}. The underlying mechanism is spectral flow of boundary-supported chiral modes under rotation. This suggests the possibility that analogous boundary-enabled vortical pumping may arise under more general boundary conditions, although establishing this lies beyond the scope of the present work.

Remarkably, the pump is independent of temperature, chemical potential, Weyl velocity, and the shape of the boundary. On the other hand, it explicitly depends on the UV cutoff $N$. Physically, $N$ corresponds to the orbital angular momentum cutoff and scales with the transverse radius of the system. As a result, \eqref{eq:charge-pump} yields a finite transport current density that is distinct from the CVE current density described in previous works focusing on bulk transport.

Thus, the present work separates the bulk and boundary contributions to vortical response. The bulk CVE is magnetization-like, while spin-polarized confinement exposes a distinct boundary transport phenomenon associated with chiral spectral flow.

\section{Quantum formulation of a rotating Weyl fermion}

We begin with general Weyl nodes of chirality $\chi=\pm1$. A minimal Hamiltonian can be written as 
\begin{align}
\label{eq:Hlab}
H_{\text{lab}} & ={\chi}v_F\mathbf{k}\cdot\boldsymbol{\sigma}-\mu.
\end{align}
We work with the common eigenbasis of operators $H_{\text{lab}}$, $J_z$, $k_\perp^2:= k_x^2+k_y^2$ and $k_z$, which preserves total angular momentum $J_z$, facilitating the analysis. In this basis, the energy spectrum is characterized by the quantum numbers $\lambda,n,k_{\perp},k_z$ ($\lambda=\pm1$)
\begin{align}
\label{eq:lab-spectrum}
    E_\text{lab}^{\lambda} &={\lambda}v_F\sqrt{k_z^2+k_\perp^2}-\mu\\
    \psi_{\lambda,n,k_{\perp},k_z} &\propto\left(\begin{array}{c}
a_{\lambda}(k_{\perp},k_{z})\mathcal{R}_n(k_{\perp}\rho)e^{in\phi}\\
b_{\lambda}(k_{\perp},k_{z})\mathcal{R}_{n+1}(k_{\perp}\rho)e^{i(n+1)\phi}
\end{array}\right)e^{ik_{z}z}\nonumber
\end{align}
where $(\rho,\phi,z):=\mathbf{r}$ are real space cylindrical coordinates, and $\mathcal{R}_n$ are Bessel functions $J_n$ with $k_\perp>0$, modified Bessel functions $I_n$ obtained by replacement $k_\perp\to i\kappa$ with $\kappa>0$, or power-law functions $\rho^{|n|}$ via the limit $k_\perp\to0$ for wave-like bulk, evanescent surface and critical states, respectively. The bulk functions are present for any boundary conditions; in contrast, the critical solutions exist only for fully spin-polarized boundary conditions, $\langle\sigma_z\rangle=\pm1$ on the boundary, while evanescent states exist only when boundary $\langle\sigma_z\rangle\neq\pm1$. As a result, the critical states may be viewed as a limiting case of the usual surface states. The explicit forms of $\mathcal{R}_n$ as well as those of the coefficients $a_{\lambda},b_{\lambda}$ are given in Sec. A1 of \cite{SM}.

\begin{figure}
    \centering
    \includegraphics[width=1.0\linewidth]{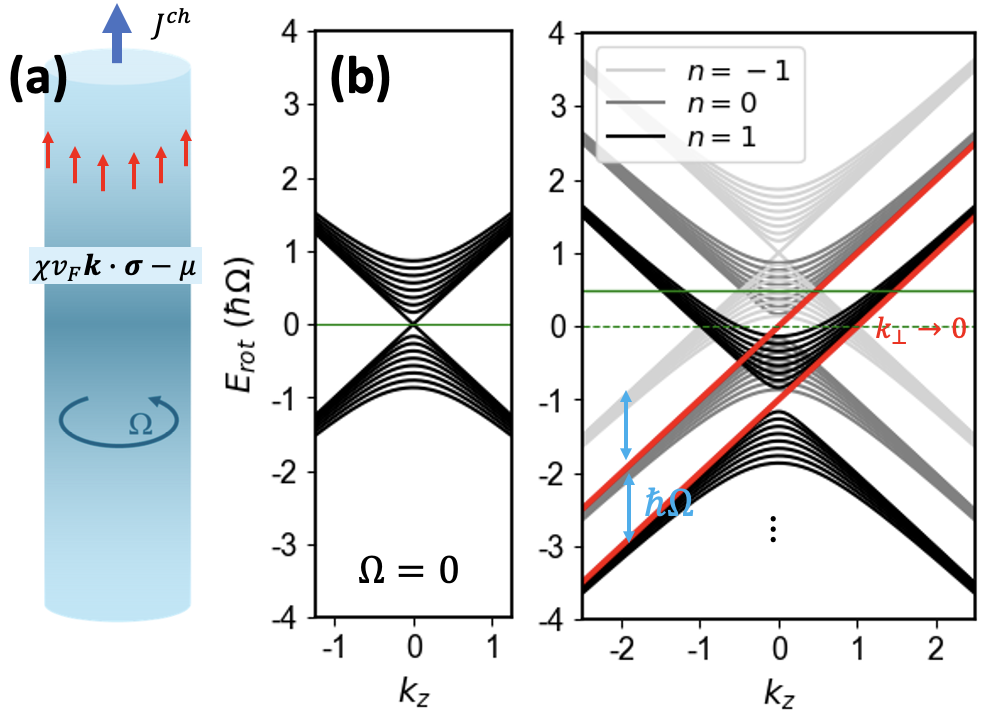}
    \caption{(a) Illustration of a rotating Weyl fermion on a cylinder with a spin-polarized boundary (red). (b) The spectrum of $H_\text{rot}$ \eqref{eq:H-rot}, consisting of subbands separated in energy by $\hbar\Omega$ (only $|n|{\le1}$ are demonstrated). For spin-polarized boundary conditions, the $n\ge0$ subbands host chiral modes (red) with $k_\perp\to0$ whose effective Fermi level changes by $(n+1/2)\hbar\Omega$ (green) under rotation, leading to quantized charge pumping. Other subbands contain only $k_\perp\neq0$ modes, which are non-chiral and lead to vanishing transport.}
    \label{fig:spectrum}
\end{figure}

To analyze rotation, we consider a frame co-rotating with angular velocity $\boldsymbol{\Omega}=\Omega\hat{\mathbf{z}}$.
In this frame, the single-particle Hamiltonian acquires a canonical coupling between the angular velocity and the total angular momentum, 
\begin{align}
\label{eq:H-rot}
H_{\text{rot}} & =H_{\text{lab}}-J_{z}\Omega
\end{align}
In the bulk, the two frames share the wavefunctions $\psi$ and quantum numbers $(n,k_{z},k_{\perp})$. The only effect of the coupling is to discretize the spectrum into angular-momentum subbands labeled by $J_z$ eigenvalues $n+1/2,n\in\mathbb{Z}$ shifting their energies as $E_{\text{rot}}^{\lambda}=E_{\text{lab}}^{\lambda}-(n+1/2)\Omega$ (Fig. \ref{fig:spectrum}). 

The above subband structure is analogous to Landau quantization underlying the CME \cite{Niemann2017, Xu2015, Cortijo2016}, with angular velocity replacing magnetic flux as the quantizing parameter and the modes dispersing along the rotation axis. Unlike the CME, however, Eq.~\eqref{eq:lab-spectrum} does not contain a bulk density of chiral modes; bulk states occur in degenerate pairs with opposite $k_z$ and opposite longitudinal velocities $\partial_{k_z}E^{\lambda}_\mathrm{lab}$. Consequently, the quantum structure of the CVE is more subtle than that of the CME. Vilenkin showed that a finite axial current nevertheless emerges on the rotation axis due to the imbalance in the occupations of up- and down-spin modes induced by rotation \cite{Vilenkin1979}. Here we further show that this local current is magnetization-like: the transport current carried by the bulk states vanishes in the thermodynamic limit despite the finite current density on the axis. Thus, the familiar semiclassical and perturbative expressions for the CVE emerge as a local manifestation of an underlying magnetization current.

\section{Bulk CVE as a magnetization current}
\label{sec:bulk-cve}

To evaluate the vortical current, we assume that the system reaches a steady state described by a Fermi-Dirac distribution in the rotating frame,
$f_{\mathrm{eq}}(H_{\mathrm{rot}})$, as is standard in semiclassical treatments of the CVE \cite{Chen2014}. The axial current density is
\begin{equation}
\label{eq:jz-general}
j_z(\mathbf{r})
=
\sum_{\lambda,n,k_z,k_\perp}
f_{\mathrm{eq}}\left(E_{\mathrm{rot}}^\lambda\right)
\psi^{*}_{\lambda,n,k_{\perp},k_z}(\mathbf{r})
\chi v_F\sigma_z
\psi_{\lambda,n,k_\perp,k_z}(\mathbf{r}),
\end{equation}
where $\chi v_F\sigma_z=\delta H_{\mathrm{lab}}/\delta k_z$ is the current operator along $\hat{\mathbf{z}}$. The sum over $k_\perp$ includes the appropriate real and imaginary values associated with bulk, surface, and critical states. Details of the calculation are given in App.~\ref{sec:j-calc}.

\begin{figure}
\includegraphics[scale=0.45]{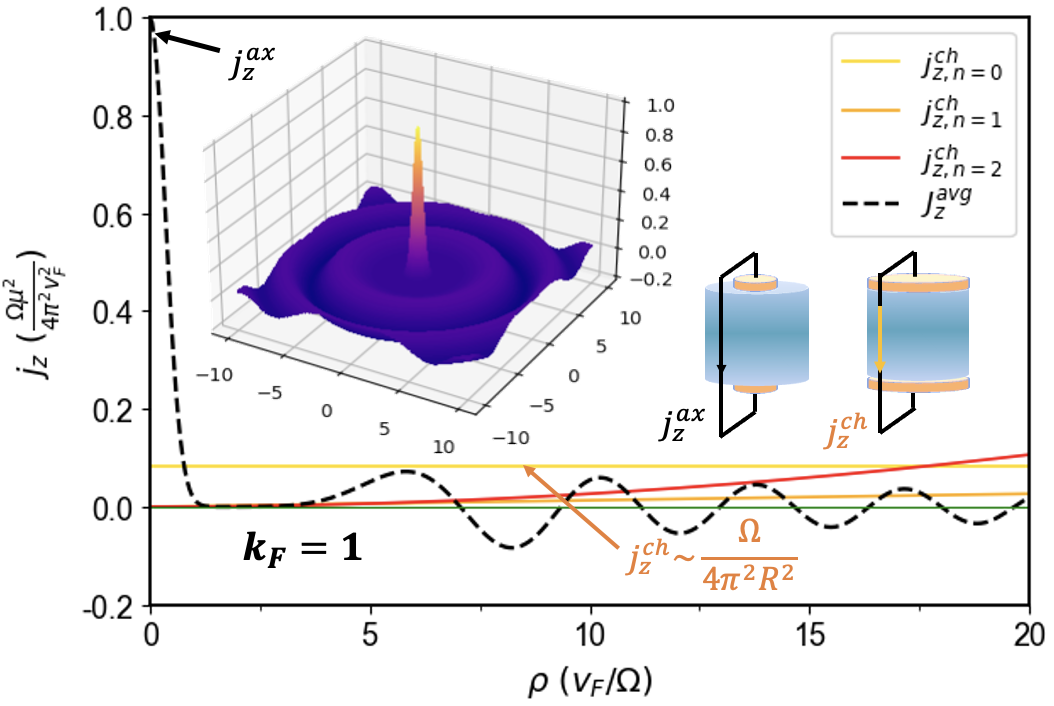}
\caption{\label{fig:jz-profile} Radial dependence of the current density $j_z$ (inset: 3D plot) and the chiral current density $j_z^{\text{ch}}$ (orange). At $\rho\to0$, the value of $j_z$, termed $j_z^\text{ax}$ here, equals the semiclassical value, while $j_z$ averages to zero over a cross-section in the thermodynamic limit due to its oscillatory nature. $N_{k_{\perp}}=11$ is adopted, and $N_{k_{\perp}}$-dependence is shown in Sec. B2 of \cite{SM}. 
}.
\end{figure}

On the rotation axis, $\rho=0$, the bulk states yield
\begin{equation}
\label{eq:jz-result}
j_{z}^{\mathrm{ax}}
=
\chi\operatorname{sgn}(\Omega)
\left[
\left(
\frac{\mu^{2}+\Omega^{2}/12}
{4\pi^{2}v_F^2}
\right)|\Omega|
+
\frac{T^{2}}{12}
\min\left(2|\mu|,|\Omega|\right)
\right].
\end{equation}
The contributions from the evanescent and critical states vanish in the thermodynamic limit. Equation~\eqref{eq:jz-result} reproduces the exact on-axis result obtained by Vilenkin \cite{Vilenkin1979}. In particular, only the $n=0,-1$ channels, corresponding to $J_z=\pm1/2$, remain finite on the axis because $J_n(0)=\delta_{n0}$. In the regime $|\Omega|\ll|\mu|$, Eq.~\eqref{eq:jz-result} reduces to the familiar semiclassical and perturbative expression for the CVE
\cite{StephanovKineticTheory,Shitade2020,Kharzeev:2016vn,
Stone2018MixedExpansion}. Away from the axis, the different
angular-momentum channels produce an oscillatory radial current
profile, as shown in Fig.~\ref{fig:jz-profile}.

The finite local current in Eq.~\eqref{eq:jz-result}, however, does not by itself imply transport through the cylinder. To determine the transport contribution, we average the bulk current over a cross-section of area $A$:
\begin{equation}
\label{eq:j-avg-definition}
j_z^{\mathrm{avg},\mathrm{bulk}}
=
\frac{1}{A}
\int_A d^2r\,
j_z^{\mathrm{bulk}}(\mathbf{r})
,
\end{equation}
where \begin{widetext}
\begin{align}
j_z^\text{bulk}(\mathbf{r}) &=  \sum_{\lambda,n,k_{\perp},k_{z}}f_{\text{eq}}\left[E_{\text{rot}}^{\lambda}(n,k_{\perp},k_z)\right]\left(\left|a_{\lambda}(k_{\perp},k_{z})\right|^{2}J_{n}^{2}(k_\perp\rho)-\left|b_{\lambda}(k_{\perp},k_{z})\right|^{2}J_{n+1}^{2}(k_\perp\rho)\right) \label{eq:j-bulk}\\
j_z^{\mathrm{avg},\mathrm{bulk}}
&=
\frac{\chi v_F}{V}
\sum_{\lambda,n,k_\perp,k_z}
f_{\mathrm{eq}}
\left(E_{\mathrm{rot}}^\lambda\right)
\left(
|a_\lambda(k_\perp,k_z)|^2
-
|b_\lambda(k_\perp,k_z)|^2
\right)
\nonumber\\
&=
\frac{\chi v_F}{V}
\sum_{\lambda,n,k_\perp,k_z}
f_{\mathrm{eq}}
\left[
\lambda v_F\sqrt{k_z^2+k_\perp^2}
-\mu
-\left(n+\frac{1}{2}\right)\Omega
\right]
\frac{\lambda k_z}
{\sqrt{k_z^2+k_\perp^2}} .
\label{eq:j-avg-algebra}
\end{align}
\end{widetext}
with $V=A\ell_z$. Here we used the fact that, in the thermodynamic limit, the two radial components have identical spatial normalizations, as discussed in Appendix \ref{subsubsec:bulk}, together with
\begin{equation}
\label{eq:spin-polarization-bulk}
|a_\lambda(k_\perp,k_z)|^2
-
|b_\lambda(k_\perp,k_z)|^2
=
\frac{\lambda k_z}
{\sqrt{k_z^2+k_\perp^2}}.
\end{equation}
The Fermi-Dirac factor in Eq.~\eqref{eq:j-avg-algebra} is even in $k_z$, whereas the final factor is odd. Contributions from states at $k_z$ and $-k_z$ therefore cancel exactly, giving
\begin{equation}
\label{eq:j-avg-bulk}
j_z^{\mathrm{avg},\mathrm{bulk}}
\xrightarrow{A\rightarrow\infty}
0.
\end{equation}

Equation~\eqref{eq:j-avg-bulk} is nonperturbative in $\Omega$ and holds for arbitrary temperature and chemical potential within the rotating frame steady state description. It extends the linear response result of Ref.~\cite{Shitade2020}, where the transport current was found to vanish to first order in $\Omega$ through a cancellation between spin vortical and Berry curvature contributions. 

It is also consistent with Bloch-type theorems forbidding a net equilibrium current density in the thermodynamic limit
\cite{Bohm1949,Ohashi1996,Yamamoto2015,Watanabe2019AModels}. In particular, the theorems apply to equilibrium systems that are suitably regularized. Although the system is out of equilibrium in the laboratory frame, the assumed steady-state distribution is an equilibrium distribution with respect to $H_{\rm rot}$. We therefore expect the same no-current constraint to be relevant to a properly regularized rotating system. A single Weyl fermion cannot itself be regularized on a compact momentum-space manifold; nevertheless, our explicit result $j_z^{\rm avg,bulk}=0$ for each Weyl fermion ensures that the bulk current also vanishes when Weyl nodes are combined in a regularized microscopic model.

The coexistence of a finite local current density and a vanishing cross-section average suggests the bulk CVE is a magnetization current density, i.e., it can be written as the curl of a well-behaved function that can be identified as a magnetization density: $\mathbf{j}^\text{bulk}(\mathbf{r})=\boldsymbol{\nabla}\times \mathbf{M}(\mathbf{r})$. The explicit form of the magnetization density is derived in Appendix \ref{subsec:cross-section average}. The result is $\mathbf{M}(\mathbf{r}) = M_\phi(\rho)\hat{\boldsymbol{\phi}}$, a purely circulating magnetization
\begin{align}
    M_\phi(\rho)&=\sum_{\lambda,n,k_{\perp},k_{z}}\frac{\chi v_F}{2k_\perp}f_{\text{eq}}\left[E_{\text{rot}}^{\lambda}(n,k_{\perp},k_z)\right]\nonumber\\
    &\qquad\times J_n(k_\perp\rho)J_{n+1}(k_\perp\rho)
\end{align}
that vanishes at both $\rho=0$ and $\rho\to\infty$.

In summary, the on-axis current is compensated by current elsewhere in the cross-section and therefore can be understood as a magnetization current. This does not preclude vortical transport in a finite system; rather, it shows that such transport cannot arise from the bulk modes alone. In the following section, we show that spin-polarized boundaries generate an additional chiral sector that supports a genuine transport current and rotation-induced charge pumping.

\section{Boundary-enforced chiral modes and vortical pumping}
\label{sec:chiral-pumping}

Having established that the bulk CVE is magnetization-like, we now turn to the additional spectral structure generated by a boundary. We consider a Weyl fermion on a geometry that is translationally invariant along $z$, with length $\ell_z$ and transverse cross-section $D$. The simplest example is a cylinder, for which $D$ is a disk. We impose the fully spin-polarized boundary condition
\begin{equation}
\label{eq:bc}
\psi(\partial D,z)
\propto
\begin{pmatrix}
1\\
0
\end{pmatrix},
\end{equation}
up to an overall phase. Such a boundary condition may arise, for example, when the Weyl system is surrounded by a ferromagnet that polarizes the boundary spin. For a circular cross-section, this condition admits no conventional surface states exponentially localized near the boundary because the relevant modified Bessel functions have no nontrivial zeros, as shown in App.~\ref{sec:j-calc}. Nevertheless, the boundary condition supports an exactly solvable family of chiral states that exists for arbitrary simply connected cross-sections.

\subsection{Exact chiral modes for a spin-polarized boundary}

We first consider the canonical Weyl node described by Eq.~\eqref{eq:Hlab}. We seek states with the chiral dispersion
\begin{equation}
\label{eq:canonical-chiral-dispersion}
E^{\rm ch}
=
\chi v_F k_z-\mu .
\end{equation}
Writing the wavefunction as
\begin{equation}
\psi(\mathbf r)
=
\begin{pmatrix}
\psi_\uparrow(\mathbf r)\\
\psi_\downarrow(\mathbf r)
\end{pmatrix},
\end{equation}
the Schr\"odinger equation reduces to
\begin{equation}
\label{eq:canonical-chiral-main}
(k_x-i k_y)\psi_\uparrow=0,
\qquad
\psi_\downarrow=0 .
\end{equation}
Introducing the complex coordinate
\begin{equation}
\zeta=x+iy,
\end{equation}
and using
\begin{equation}
k_x-i k_y
=
-i(\partial_x-i\partial_y)
=
-2i\partial_{\bar\zeta},
\end{equation}
Eq.~\eqref{eq:canonical-chiral-main} becomes
\begin{equation}
\partial_{\bar\zeta}\psi_\uparrow=0.
\end{equation}
Thus, the transverse wavefunction is holomorphic. For an arbitrary simply connected cross-section $D$, every regular holomorphic function $f(\zeta)$ therefore produces a chiral eigenstate,
\begin{equation}
\label{eq:general-holomorphic-main}
\psi^{\rm ch}(\mathbf r)
=
\frac{e^{ik_z z}}{\sqrt{\ell_z}}
\begin{pmatrix}
f(\zeta)\\
0
\end{pmatrix},
\qquad
\partial_{\bar\zeta}f=0,
\end{equation}
with dispersion \eqref{eq:canonical-chiral-dispersion}.

The polarized boundary condition also fixes this family uniquely. The opposite-spin component obeys the conjugate analytic equation and vanishes on the entire boundary. By uniqueness, it must therefore vanish throughout $D$. Hence every chiral mode compatible with Eq.~\eqref{eq:bc} is of the form \eqref{eq:general-holomorphic-main}.

For a circular cross-section of radius $R$, a convenient orthonormal basis is provided by the monomials $f_n(\zeta)\propto\zeta^n$, yielding
\begin{equation}
\label{eq:chiral-modes-main}
\psi_{n,k_z}^{\rm ch}(\rho,\phi,z)
=
\frac{e^{ik_z z}}{R^{n+1}}
\sqrt{\frac{n+1}{\pi\ell_z}}
\begin{pmatrix}
\rho^n e^{in\phi}\\
0
\end{pmatrix},
n=0,1,2,\ldots .
\end{equation}
Each nonnegative integer $n$ therefore produces one spin-polarized chiral branch with the same dispersion \eqref{eq:canonical-chiral-dispersion}. These states may be viewed as the $k_\perp\rightarrow0$ limits of the bulk modes built from Bessel functions. In particular, the $n=0$ member is spatially uniform in the transverse plane:
\begin{equation}
\psi_{0,k_z}^{\rm ch}(\mathbf r)
=
\frac{e^{ik_z z}}{\sqrt{A\ell_z}}
\begin{pmatrix}
1\\
0
\end{pmatrix}.
\end{equation}

For the opposite boundary polarization,
\begin{equation}
\psi(\partial D,z)
\propto
\begin{pmatrix}
0\\
1
\end{pmatrix},
\end{equation}
the corresponding modes are anti-holomorphic,
\begin{equation}
\psi_\uparrow=0,
\qquad
\psi_\downarrow\propto\bar{\zeta}^{\,n}e^{ik_z z},
\end{equation}
and disperse as
\begin{equation}
E^{\rm ch}
=
-\chi v_F k_z-\mu .
\end{equation}

\subsection{Rotation-induced spectral flow}

We now return to the canonical Weyl node and show that the chiral branches support a genuine transport current under rotation. Let $s=\pm1$ denote the boundary spin polarization. In the rotating frame, the energy of branch $n$ shifts according to
\begin{equation}
\label{eq:chiral-rot-energy}
E_{\rm rot}^{\rm ch}(n,s)
=
E_{\rm lab}^{\rm ch}(s)
-
s\left(n+\frac{1}{2}\right)\Omega .
\end{equation}
Rotation therefore changes the occupation of each chiral branch relative to the laboratory frame.

The rotation-induced current carried by branch $n$ is
\begin{align}
I_n^{\rm ch}
={}&
s\chi v_F
\int\frac{dk_z}{2\pi}
\Bigg[
f_{\rm eq}
\left(
E_{\rm lab}^{\rm ch}
-
s\left(n+\frac{1}{2}\right)\Omega
\right)
\nonumber\\
&\hspace{4.3cm}
-
f_{\rm eq}
\left(
E_{\rm lab}^{\rm ch}
\right)
\Bigg].
\label{eq:single-branch-current}
\end{align}
Changing variables from $k_z$ to $E_{\rm lab}^{\rm ch}$ and using
\begin{equation}
\int_{-\infty}^{\infty}dE\,
\left[
f(E-\epsilon)-f(E)
\right]
=
\epsilon,
\end{equation}
gives
\begin{equation}
\label{eq:single-branch-result}
I_n^{\rm ch}
=
\frac{\chi\Omega}{2\pi}
\left(n+\frac{1}{2}\right).
\end{equation}
The result is independent of temperature, chemical potential, boundary polarization, and the Weyl velocity. Summing over the $N$ chiral branches retained by the microscopic theory gives
\begin{align}
I^{\rm ch}
&=
\sum_{n=0}^{N-1}I_n^{\rm ch}
\nonumber\\
&=
\frac{\chi\Omega}{2\pi}
\sum_{n=0}^{N-1}
\left(n+\frac{1}{2}\right)
\nonumber\\
&=
\chi\frac{\Omega}{4\pi}N^2 .
\label{eq:I-pump}
\end{align}
Integrating over time yields the axial charge transported between the two ends of the system,
\begin{equation}
\label{eq:charge-pump-main}
\Delta Q^{\rm ch}
=
\int dt\,I^{\rm ch}
=
\chi\frac{\Delta\theta}{4\pi}N^2.
\end{equation}
Thus, for fixed $N$, a global rotation by $4\pi$ transfers the integer charge $\chi N^2$ along the rotation axis.

The continuum Weyl Hamiltonian formally produces an unbounded ladder of angular-momentum channels. In a microscopic system, however, this ladder terminates once the transverse variation of the wavefunction reaches the lattice or boundary scale. For a cylinder of radius $R$ and microscopic cutoff $a$, one expects parametrically
\begin{equation}
N\sim\frac{R}{a}.
\end{equation}
Consequently, $I^{\rm ch}\propto R^2$, and the corresponding current density remains finite in the thermodynamic limit. The magnitude of the pump is therefore UV-sensitive and nonuniversal, even though the spectral-flow relation for each branch, Eq.~\eqref{eq:single-branch-result}, is independent of temperature, chemical potential, and band velocity.

This boundary-enforced current is qualitatively distinct from the bulk CVE discussed in Sec.~\ref{sec:bulk-cve}. The bulk contribution has a finite local current density but a vanishing cross-section average, whereas the chiral modes produce a nonzero transport current through the sample.

\subsection{Generality of the chiral sector}

The existence of exactly solvable chiral modes is not restricted to a spin-$1/2$ Weyl node of unit monopole charge. Consider the generalized Hamiltonian
\begin{equation}
H_{\rm lab}
=
2\chi
\left(
v k_zS_z
+
u k_-^{\,q}S_+
+
u k_+^{\,q}S_-
\right)
-\mu,
\end{equation}
describing a Weyl node with spin $S$ and monopole charge $q$. If the boundary selects the extremal spin state $|S\rangle$, then $S_+|S\rangle=0$, and chiral states may be written in factorized form,
\begin{equation}
\psi^{\rm ch}(\mathbf r)
=
\frac{e^{ik_z z}}{\sqrt{\ell_z}}
w(\zeta,\bar\zeta)|S\rangle .
\end{equation}
The transverse profile obeys
\begin{equation}
k_-^{\,q}w(\zeta,\bar\zeta)=0,
\end{equation}
and therefore admits the polyanalytic family
\begin{equation}
w(\zeta,\bar\zeta)
\propto
\bar\zeta^{\,m}\zeta^n,
\qquad
0\leq m<q.
\end{equation}
Each such profile produces a chiral branch with dispersion
\begin{equation}
E^{\rm ch}
=
2S\chi v k_z-\mu.
\end{equation}
Because this construction depends only on the ladder-operator algebra and the extremally polarized boundary condition, it applies to arbitrary cross-sectional shapes. For $q>1$, however, the polyanalytic family need not exhaust all possible chiral modes. The detailed counting and the resulting pumping contribution for arbitrary $S$ and $q$ are given in App.~\ref{sec:chiral-topology}.

The fully spin-polarized boundary therefore provides an exactly solvable setting in which rotation drives spectral flow through a boundary-enforced chiral sector. Whether an analogous pump persists for generic boundary conditions, where these modes may hybridize with conventional surface states, remains an open question.

\section{Conclusions}

We have studied the chiral vortical effect for a Weyl fermion in a finite geometry using an exact quantum treatment. The formalism reproduces the exact on-axis current previously obtained by Vilenkin while showing that the spatially averaged bulk current vanishes in the thermodynamic limit. The bulk CVE is therefore magnetization-like: it produces a finite local current density without net charge transport through the sample.

We further showed that fully spin-polarized boundaries support an exactly solvable family of chiral modes that is distinct from the bulk spectrum. Rotation shifts these modes according to their angular momentum, producing boundary-induced spectral flow and a transport current that pumps axial charge between the two ends of the system. Although the magnitude of the pump depends on the ultraviolet cutoff through the number of chiral modes, its mechanism is remarkably insensitive to temperature, chemical potential, Weyl velocity, and the shape of the boundary.

Our results separate the bulk and boundary contributions to vortical response in finite Weyl systems. The bulk response is governed by equilibrium magnetization currents, while genuine transport arises from a boundary-enforced chiral sector. 

From an experimental perspective, observing these effects in Weyl semimetals is likely to be challenging. The relevant angular velocity is that of the electronic fluid relative to the lattice rather than a directly controllable laboratory rotation, and in realistic transport settings it is itself induced by external perturbations such as electric fields or temperature gradients. Consequently, isolating the vortical contribution from other transport mechanisms will require careful analysis. Moreover, the magnitude of the boundary pump depends on the microscopic cutoff through the number of chiral modes, making it material dependent. Finally, Weyl semimetals host multiple Weyl nodes with net zero chirality, so the contributions from different nodes are expected to cancel, consistent with the absence of a net equilibrium current in a regularized system. On the other hand, engineered platforms with greater control over rotation and confinement along with an ability to measure momentum-resolved currents, such as via time-of-flight experiments with ultracold atomic systems, may provide a cleaner route to exploring the quantum structure of vortical response described here.

\begin{acknowledgements}
    We thank Anton Burkov, Shun-Chiao Chang, and Carlos Ordonez for useful discussions. This work was supported by the Department of Energy grant number DE-SC0022264.
\end{acknowledgements}

\appendix

\begin{widetext}

\section{Weyl fermion on a cylinder}

\subsection{Hamiltonians and spectrum}
\label{subsec:Hamiltonian-and-spectrum}

In this appendix we present a self-contained derivation of the exact spectrum of a Weyl fermion confined to a cylindrical geometry. While Sec. II summarizes the key ingredients used in the main text, the complete wavefunctions, normalization, and boundary-dependent spectrum are collected here for reference.

The Hamiltonian for a Weyl fermion of chirality $\chi$ can be written
as ($\hbar=1$) 
\begin{align}
H_{\text{lab}} & =\chi{v_F}\mathbf{k}\cdot\boldsymbol{\sigma}-\mu\\
 & =\chi\left(\begin{array}{cc}
-i\partial_{z} & -ie^{-i\phi}\left(\partial_{r}-\frac{i}{\rho}\partial_{\phi}\right)\\
-ie^{i\phi}\left(\partial_{r}+\frac{i}{\rho}\partial_{\phi}\right) & i\partial_{z}
\end{array}\right)-\mu
\end{align}
where $(\rho,\phi,z)$ are real-space cylindrical co-ordinates. Besides
$(k_{x},k_{y},k_{z})$, $H_{\text{lab}}$ also preserves the quantum
numbers $k_{z}$, $k_{\perp}^{2}=k_{x}^{2}+k_{y}^{2}$, and $J_{z}=L_{z}+\sigma_{z}/2\equiv n+1/2$,
which will be more convenient for analyzing the rotating system. The
eigenstates are of the form: 
\begin{equation}
\psi_{\lambda,n,k_{\perp},k_{z}}(\rho,\phi,z)=\frac{1}{\sqrt{N}}\left(\begin{array}{c}
a_{\lambda}(k_{\perp},k_{z})\mathcal{R}_n(k_{\perp}\rho)e^{in\phi}\\
b_{\lambda}(k_{\perp},k_{z})\mathcal{R}_{n+1}(k_{\perp}\rho)e^{i(n+1)\phi}
\end{array}\right)e^{ik_{z}z}
\end{equation}
where $\mathcal{R}_{n,n+1}$ are radial wavefunctions to be determined. In this basis, $H_{\text{lab}}$ transforms to 
\begin{equation}
H_{\text{lab}}=\chi{v_F}\left(\begin{array}{cc}
k_{z} & -i\left(\partial_{r}+\frac{n+1}{\rho}\right)\\
-i\left(\partial_{r}-\frac{n}{\rho}\right) & -k_{z}
\end{array}\right)-\mu
\end{equation}

In this work, we restrict to spin-polarized boundary conditions:
\begin{equation}
\label{eq:polarized}
\sigma_z\psi(R,\phi,z) = s\psi(R,\phi,z)\quad,\quad s=\pm1
\end{equation}

When the Weyl fermion is subject to rotation, the Hamiltonian in the
rotating frame contains a coupling between the angular velocity and
a suitable angular momentum. In the literature, two choices have been
widely adopted, where the difference is the exclusion or inclusion
of spin-vorticity coupling, or equivalently, coupling of angular velocity
to spatial or total angular momentum. We restrict to the latter: 
\begin{align}
H_{\text{rot}} & =H_{\text{lab}}-\mathbf{J}\cdot\boldsymbol{\Omega}
\end{align}
where $\mathbf{J}=\mathbf{L}+\frac{\boldsymbol{\sigma}}{2}$. Taking $\boldsymbol{\Omega}=\Omega\hat{\mathbf{z}}$, the energies change as
\begin{align}
\label{eq:E-rot-E-lab}
E_\text{rot} = E_\text{lab}-\left(n+\frac{1}{2}\right)\Omega\quad,\quad n\in\mathbb{Z}
\end{align}
while the eigenstates remain unchanged. Below, we write down the explicit forms of the eigenstates.

\subsubsection{Bulk states}
\label{subsubsec:bulk}
First, we describe oscillating bulk states defined by the Bessel functions $J_n$ of the first kind, which satisfy the recurrence relations: 
\begin{align}
\left(\partial_{\rho}+\frac{n+1}{\rho}\right)J_{n+1}(k_{\perp}\rho) & =k_{\perp}J_{n}(k_{\perp}\rho)\\
\left(\partial_{\rho}-\frac{n}{\rho}\right)J_{n}(k_{\perp}\rho) & =-k_{\perp}J_{n+1}(k_{\perp}\rho)
\end{align}
Wavefunctions based on the second-kind Bessel functions $Y_{n}$ are trivially excluded as they diverge at $\rho\to0$. This yields a bulk spectrum in terms of the quantum numbers $(k_{z},k_{\perp},n)$,
\begin{align}
E_{\text{lab}}^{\lambda}(k_{z},k_{\perp},n) & =\lambda\sqrt{k_{z}^{2}+k_{\perp}^{2}}-\mu:= E_{\text{lab}}^{\lambda}(k_{z},k_{\perp})\\
\psi_{\lambda,n,k_{\perp},k_{z}}(\rho,\phi,z) & =\frac{1}{\sqrt{N_{n}}}\left(\begin{array}{c}
a_{\lambda}(k_{\perp},k_{z})J_{n}(k_{\perp}\rho)e^{in\phi}\\
b_{\lambda}(k_{\perp},k_{z})J_{n+1}(k_{\perp}\rho)e^{i(n+1)\phi}
\end{array}\right)e^{ik_{z}z}
\end{align}
where
\begin{equation}
\left(\begin{array}{c}
a_{\lambda}(k_{\perp},k_{z})\\
b_{\lambda}(k_{\perp},k_{z})
\end{array}\right)=\frac{1}{\sqrt{2\sqrt{k_{z}^{2}+k_{\perp}^{2}}}\sqrt{\sqrt{k_{\perp}^{2}+k_{z}^{2}}-\lambda k_{z}}}\left(\begin{array}{c}
ik_{\perp}\\
k_{z}-\lambda\sqrt{k_{z}^{2}+k_{\perp}^{2}}
\end{array}\right)
\end{equation}
and $k_\perp$ takes discrete values from an $n$-dependent set defined by the roots of $J_n(k_\perp R)$ or $J_{n+1}(k_\perp R)$ for $s=\mp1$, respectively, following \eqref{eq:polarized}. Note that the lab frame energies are independent of $n$ while rotating frame energies develop subbands according to \eqref{eq:E-rot-E-lab}.

To determine the normalization constant $N_n$, we focus on strictly wave-like states with $k_{\perp}>0$. States with
$k_{\perp}=0$ will be considered shortly in Sec. \ref{subsubsec:chiral-modes}.
Denoting the non-zero values by $k_{\perp,n,i}$, quantized due to
appropriate boundary conditions, we must have 
\begin{align}
\intop d^{3}r\left|\psi_{\lambda,n,k_{\perp},k_{z}}(\rho,\phi,z)\right|^{2}=\frac{1}{N_{n}}2\pi\ell_{z}\left[\left|a_{\lambda}(k_{\perp,n},k_{z})\right|^{2}\intop_{0}^{R}\rho d\rho J_{n}^{2}(k_{\perp,n,i}\rho)+\left|b_{\lambda}(k_{\perp,n},k_{z})\right|^{2}\intop_{0}^{R}\rho d\rho J_{n+1}^{2}(k_{\perp,n,i}\rho)\right] & =1
\end{align}
on a cylinder of radius $R$. A standard Bessel function identity
is 
\begin{align}
\intop_{0}^{\mathcal{X}}xdxJ_{n}^{2}(x) & =\frac{\mathcal{X}^{2}}{2}\left[J_{n}^{2}(\mathcal{X})-J_{n+1}(\mathcal{X})J_{n-1}(\mathcal{X})\right]\approx\frac{\mathcal{X}}{\pi}
\end{align}
independent of $n$, assuming the asymptotic form, $J_{n}(x)\approx\sqrt{\frac{2}{\pi x}}\cos\left(x-n\pi/2-\pi/4\right)$
for $x\gg|n|$. This gives the normalization 
\begin{equation}
N_{n}=\frac{2\ell_{z}R}{k_{\perp}}
\end{equation}
and the wavefunction can be compactly written as 
\begin{equation}
\psi_{\lambda,n,k_{\perp},k_{z}}(\rho,\phi,z)=\sqrt{\frac{k_{\perp}}{2\ell_{z}R}}\left(\begin{array}{c}
a_{\lambda}(k_{\perp},k_{z})J_{n}(k_{\perp}\rho)e^{in\phi}\\
b_{\lambda}(k_{\perp},k_{z})J_{n+1}(k_{\perp}\rho)e^{i(n+1)\phi}
\end{array}\right)e^{ik_{z}z}
\end{equation}
where $\ell_{z}$ is the length of the cylinder.

\subsubsection{Chiral modes}
\label{subsubsec:chiral-modes}

Besides wave-like bulk and evanescent surface states, the spectrum of $H_\text{lab}$ or $H_\text{rot}$ on a cylinder contains another set of states. Section IV introduces the chiral modes from a boundary perspective and emphasizes their physical consequences. Here we derive their explicit wavefunctions and normalization for the cylindrical geometry. These states can be obtained as the $k_\perp\to0$ limit of the bulk states, but are more easily written down directly. Explicitly, they are:

\begin{align}
\label{eq:chiral-modes}
    E^\text{ch}_{\text{lab},\uparrow} = \chi v_F k_z-\mu\quad &,\quad
    \psi_{n,k_z,\uparrow}^\text{ch}(\rho,\phi,z) = \frac{1}{R^{n+1}}\sqrt{\frac{n+1}{\pi\ell_z}}\left(
    \begin{array}{c}
        \rho^ne^{in\phi} \\
        0
    \end{array}
    \right)e^{ik_zz}\\
    E^\text{ch}_{\text{lab},\downarrow} = -\chi v_F k_z-\mu\quad &,\quad
     \psi_{n,k_z,\downarrow}^\text{ch}(\rho,\phi,z) = \frac{1}{R^{n+1}}\sqrt{\frac{n+1}{\pi\ell_z}}\left(
    \begin{array}{c}
        0 \\
        \rho^ne^{-in\phi} \\
    \end{array}
    \right)e^{ik_zz}
\end{align}
for $n\geq0$, as can be verified by direct substitution. These states are clearly spin-polarized and chiral. For a given spin-polarized boundary condition \eqref{eq:polarized}, either $\psi^\text{ch}_{n,k_z,\uparrow}$ or $\psi^\text{ch}_{n,k_z,\downarrow}$ exist, while more general boundary conditions disallow both sets of states.

Under rotation, the energies change as
\begin{equation}
    E^\text{ch}_{\text{rot},s} = E^\text{ch}_{\text{lab},s} - s\left(n+\frac{1}{2}\right)\Omega
\end{equation}
where $s=\pm1$ corresponds to $\uparrow/\downarrow$. Note the extra factor of $s$ above compared to \eqref{eq:E-rot-E-lab}, which reflects the effect of the boundary. In App. \ref{sec:j-calc}, we will repeatedly use the identity,
\begin{align}
    s\chi v_F\intop_{-\infty}^\infty \frac{dk_z}{2\pi}\left[f_\text{eq}(E^\text{ch}_{\text{rot},s})-f_\text{eq}(E^\text{ch}_{\text{lab},s})\right] 
    &=     \intop_{-\infty}^\infty \frac{dk_z}{2\pi}\frac{dE^\text{ch}_{\text{lab},s}}{dk_z}\left[f_\text{eq}(E^\text{ch}_{\text{lab},s}-s(n+1/2)\Omega)-f_\text{eq}(E^\text{ch}_{\text{lab},s})\right] \\
    &= \frac{s\chi}{2\pi}\intop_{-\infty}^\infty dE \left[f_\text{eq}(E-s(n+1/2)\Omega)-f_\text{eq}(E)\right]\\
    &= \frac{\chi}{2\pi}\left(n+\frac{1}{2}\right)\Omega
\end{align}
where $f_\text{eq}$ is the equilibrium Fermi-Dirac distribution.

\subsubsection{Surface states}

Surface states can be derived by invoking recurrence relations for the modified Bessel functions $I_n$ of the first kind: 
\begin{align}
\left(\partial_{r}+\frac{n+1}{\rho}\right)I_{n+1}(\kappa\rho) & =\kappa I_{n}(\kappa\rho)\\
\left(\partial_{r}-\frac{n}{\rho}\right)I_{n}(\kappa\rho) & =\kappa I_{n+1}(\kappa\rho)
\end{align}
This yields the spectrum
\begin{align}
\mathcal{E}_{\text{lab}}^{\lambda}v_F(k_{z},\kappa) & =\lambda\sqrt{k_{z}^{2}-\kappa^{2}}-\mu\\
\psi_{\lambda,n,\kappa,k_{z}}^{\text{sf}}(\rho,\phi,z) & =\frac{1}{\sqrt{N_n}}\left(\begin{array}{c}
a_{\lambda}^{\text{sf}}(\kappa,k_{z})I_{n}(\kappa\rho)e^{in\phi}\\
b_{\lambda}^{\text{sf}}(\kappa,k_{z})I_{n+1}(\kappa\rho)e^{i(n+1)\phi}
\end{array}\right)e^{ik_{z}z}
\end{align}
with
\begin{equation}
\left(\begin{array}{c}
a_{\lambda}^{\text{sf}}(\kappa,k_{z})\\
b_{\lambda}^{\text{sf}}(\kappa,k_{z})
\end{array}\right)=\frac{1}{\sqrt{2k_{z}\left(k_{z}-\lambda\sqrt{k_{z}^{2}-\kappa^{2}}\right)}}\left(\begin{array}{c}
i\kappa\\
k_{z}-\lambda\sqrt{k_{z}^{2}-\kappa^{2}}
\end{array}\right)
\end{equation}

Once again, the energies are independent of $n$ and wavefunctions based on the modified Bessel functions of the second kind, $K_n$ are excluded as they diverge at the origin.

Normalization of surface states follows from the identity
\begin{equation}
\intop_{0}^{\mathcal{X}}xdxI_{n}^{2}(x)=\frac{\mathcal{X}^{2}}{2}\left[I_{n}^{2}(\mathcal{X})-I_{n+1}(\mathcal{X})I_{n-1}(\mathcal{X})\right]\approx\frac{\mathcal{X}^{2}}{2}\frac{e^{2\mathcal{X}}}{2\pi\mathcal{X}^{2}}=\frac{e^{2\mathcal{X}}}{4\pi}
\end{equation}
independent of $n$ for $\kappa R\gg n^2$ using the asymptotic form 
\begin{equation}
I_{n}(x)\approx\frac{e^{x}}{\sqrt{2\pi x}}\left[1-\frac{4n^{2}-1}{8x}+\dots\right];x\gg|n^{2}-1/4|
\end{equation}
This yields the normalization constant 
\begin{equation}
N_{n}=\frac{\ell_{z}e^{2\mathcal{\kappa}R}}{2\kappa^{2}}
\end{equation}
and normalized surface states
\begin{equation}
\label{eq:surface-states}
\psi_{\lambda,n,\kappa,k_{z}}^{\text{sf}}(\rho,\phi,z)=\kappa e^{-\kappa R}\sqrt{\frac{2}{\ell_{z}}}\left(\begin{array}{c}
a_{\lambda}^{\text{sf}}(\kappa,k_{z})I_{n}(\kappa\rho)e^{in\phi}\\
b_{\lambda}^{\text{sf}}(\kappa,k_{z})I_{n+1}(\kappa\rho)e^{i(n+1)\phi}
\end{array}\right)e^{ik_{z}z}
\end{equation}

However, boundary conditions (\ref{eq:polarized}) lead to a surprising
consequence -- the absence of surface states. This is because (\ref{eq:polarized})
requires either $I_{n}(\kappa R)$ or $I_{n+1}(\kappa R)$ to vanish,
but these Bessel functions vanish only if $\kappa R=0$ whereas we have assumed $\kappa\neq0$ at the outset. Thus, for spin-polarized
boundary conditions, there are no conventional exponentially localized surface states. On the other hand, the fact that the chiral modes appear only for spin-polarized boundary conditions suggests that they correspond to a certain critical limit of true surface states achieved by the special boundary conditions.

\section{Rotation-induced current densities}\label{sec:j-calc}

In this appendix we provide the detailed calculations underlying Sec. III. For completeness, we derive both the local current on the rotation axis and the cross-section-averaged current directly from the exact quantum eigenstates before evaluating the contributions of the boundary chiral modes.

Our basic premise is that the system acquires an equilibrium Fermi-Dirac
distribution $f_{\text{eq}}$ in the rotating frame: 
\begin{equation}
\hat{\rho}=f_{\text{eq}}(H_{\text{rot}})
\end{equation}
Using $\hat{\rho}$, we will write down the current density $j_{z}(\rho,\phi,z)$
and evaluate it on the axis $(\rho=0)$, on the cylinder surface $(\rho=R)$
and averaged over the cylindrical cross-section. We restrict to polarized
boundary conditions (\ref{eq:polarized}) which results in the absence
of surface states. Therefore, it suffices to calculate current densities
only using the bulk states. Importantly, an intensive quantity like
the current density should not depend on the boundary conditions,
so we expect the results in this section to hold for arbitrary boundary
conditions.

The current density along $z$ is 
\begin{align}
j_{z}(\rho,\phi,z) & =\sum_{n,k_{\perp},k_{z}}\chi\text{tr}\left[\sigma_{z}\hat{\rho}(\rho,\phi,z)\right]\\
 & =\sum_{\lambda,n,k_{\perp},k_{z}}f_{\text{eq}}\left(E_{\text{rot}}^{\lambda}\right)\psi_{\lambda,n,k_{\perp},k_{z}}^*(\rho,\phi,z)v_F\sigma_{z}\psi_{\lambda,n,k_{\perp},k_{z}}(\rho,\phi,z)\\
 & =\sum_{\lambda,n}\frac{\ell_{z}}{2\pi}\int dk_{z}\frac{R}{\pi}\int dk_{\perp}f_{\text{eq}}\left(E_{\text{rot}}^{\lambda}\right)\psi_{\lambda,n,k_{\perp},k_{z}}^*(\rho,\phi,z)v_F\sigma_{z}\psi_{\lambda,n,k_{\perp},k_{z}}(\rho,\phi,z)
\end{align} 
Due to translational symmetry along $z$ and rotational symmetry about
the $z$ axis, $j_{z}(\rho,\phi,z)$ is expected to depend only on
$\rho$, and will be denoted $j_{z}(\rho)$ henceforth.

\subsection{On the axis\label{subsec:rho0 current}}

\subsubsection{From bulk states}

We first revisit the on-axis current originally derived by Vilenkin \cite{Vilenkin1979}. The calculation is reproduced here because it provides the starting point for the subsequent analysis of the cross-section-averaged current and facilitates comparison with the boundary contribution.

At $\rho=0$, i.e., on the rotation axis, the current due to the bulk, wave-like states is 
\begin{align}
j_z^\text{ax,bulk}=\chi{v_F}\sum_{\lambda,n}\intop_{\mathbf{k}}f_{\text{eq}}\left[E_{\text{rot}}^{\lambda}(n,k_{\perp},k_{z})\right]\left(\left|a_{\lambda}(k_{\perp},k_{z})\right|^{2}J_{n}^{2}(0)-\left|b_{\lambda}(k_{\perp},k_{z})\right|^{2}J_{n+1}^{2}(0)\right)\label{eq:jz(0)-step-2}
\end{align}
where $\intop_{\mathbf{k}}=\int\frac{dk_{z}k_{\perp}dk_{\perp}}{(2\pi)^{2}}$. Note that 
\begin{align}
\left|a_{\lambda}(k_{\perp},k_{z})\right|^{2} & =\left\langle \psi_{\lambda,n,k_{\perp},k_{z}}\left|\frac{1+\sigma_{z}}{2}\right|\psi_{\lambda,n,k_{\perp},k_{z}}\right\rangle =\frac{1}{2}\left(1+\frac{\lambda k_{z}}{\sqrt{k_{z}^{2}+k_{\perp}^2}}\right)\nonumber \\
\left|b_{\lambda}(k_{\perp},k_{z})\right|^{2} & =\left\langle \psi_{\lambda,n,k_{\perp},k_{z}}\left|\frac{1-\sigma_{z}}{2}\right|\psi_{\lambda,n,k_{\perp}^{2},k_{z}}\right\rangle =\frac{1}{2}\left(1-\frac{\lambda k_{z}}{\sqrt{k_{z}^{2}+k_{\perp}^{2}}}\right)\label{eq:B7}
\end{align}
Since $J_{n}(0)=\delta_{n0}$, the only terms that survive the summation
over $n$ in (\ref{eq:jz(0)-step-2}) are $n=0$ and $n=-1$. Thus,
\begin{align}
j_z^\text{ax,bulk} & =\chi{v_F}\sum_{\lambda}\intop_{\mathbf{k}}f_{\text{eq}}\left[E_{\text{rot}}^{\lambda}(k_{z},k_{\perp},0)\right]\left|a_{\lambda}(k_{\perp},k_{z})\right|^{2}-f_{\text{eq}}\left[E_{\text{rot}}^{\lambda}(k_{z},k_{\perp},-1)\right]\left|b_{\lambda}(k_{\perp},k_{z})\right|^{2}\nonumber \\
 & =\frac{\chi}{2}v_F\sum_{\lambda}\intop_{\mathbf{k}}\left\{ f_{\text{eq}}\left[E_{\text{rot}}^{\lambda}(k_{z},k_{\perp},0)\right]-f_{\text{eq}}\left[E_{\text{rot}}^{\lambda}(k_{z},k_{\perp},-1)\right]\right\} \nonumber \\
 & +\frac{\chi}{2}v_F\sum_{\lambda}\intop_{\mathbf{k}}\left\{ f_{\text{eq}}\left[E_{\text{rot}}^{\lambda}(k_{z},k_{\perp},0)\right]+f_{\text{eq}}\left[E_{\text{rot}}^{\lambda}(k_{z},k_{\perp},-1)\right]\right\} \frac{\lambda k_{z}}{\sqrt{k_{z}^{2}+k_{\perp}^{2}}}
\end{align}
The second term above is odd in $k_{z}$ and vanishes, leaving behind
\begin{equation}
\label{eq:jz-ax-last-integral}
j_z^\text{ax,bulk}=\frac{\chi}{2}v_F\sum_{\lambda}\intop_{\mathbf{k}}\left\{ f_{\text{eq}}\left[\lambda{v_F}\sqrt{k_{z}^{2}+k_{\perp}^{2}}-\mu-\frac{1}{2}\Omega\right]-f_{\text{eq}}\left[\lambda{v_F}\sqrt{k_{z}^{2}+k_{\perp}^{2}}-\mu+\frac{1}{2}\Omega\right]\right\} 
\end{equation}
At this stage, the connection with semiclassical treatments becomes transparent, reproducing the observation originally made by Vilenkin \cite{Vilenkin1979}. The exact quantum expression reduces to the standard semiclassical phase-space integral after expanding in $\Omega$. The integrand in $j_z^{\rm ax}$ depends only on the
rotationally invariant combination $k_z^2+k_\perp^2$ and contains no
residual dependence on the angular-momentum quantum number $n$, which
has been eliminated by evaluating the current on the rotation axis.
Writing $\intop_{\mathbf{k}}=\int dk_z\,k_\perp dk_\perp/(2\pi)^2$ as
$\int d^3k/(2\pi)^3$ after trivial angular averaging in the transverse
plane, one recovers precisely the semiclassical phase-space expression
for the chiral vortical current density \cite{StephanovKineticTheory,Chen2014} upon Taylor-expanding in $\Omega$. Thus, evaluating the exact quantum current density at $\rho=0$ implements the same projection as the usual semiclassical limit: it removes sensitivity to angular-momentum quantization and quantum interference, leaving a smooth phase-space integral.

On the other hand, the present quantum treatment yields exact expressions in $\Omega$ without Taylor expansion. Specifically, at low $T$, Sommerfeld expansion simplifies \eqref{eq:jz-ax-last-integral} to
\begin{align}
j_z^\text{ax,bulk} & =\frac{\chi}{2}\frac{1}{(2\pi)^{3}v_F^2}\left\{ V_{|\mu+\Omega/2|}-V_{|\mu-\Omega/2|}\right\} +\frac{\chi}{2}\frac{1}{(2\pi)^{3}}\frac{\pi^{2}T^{2}}{6}\left\{ S_{|\mu+\Omega/2|}^{\prime}-S_{|\mu-\Omega/2|}^{\prime}\right\} 
\end{align}
where $V_{a}$ and $S_{a}$ are the volume and surface area of a sphere
of radius $a$. Explicitly, 
\begin{align}
j_z^\text{ax,bulk} & =\frac{\chi}{3}\frac{|\mu+\Omega/2|^{3}-|\mu-\Omega/2|^{3}}{(2\pi)^{2}v_F^2}+\chi\frac{1}{(2\pi)^{3}}\frac{\pi^{2}T^{2}}{6}8\pi\left\{ |\mu+\Omega/2|-|\mu-\Omega/2|\right\} \nonumber \\
 & =\chi\left(\frac{\mu^{2}+\Omega^{2}/12}{4\pi^{2}v_F^2}\right)\Omega+\frac{T^{2}}{6}\text{sgn}(\Omega)\min\left(|\mu|,\frac{|\Omega|}{2}\right)
\end{align}
The well-known result for the CVE corresponds to the $|\mu|>|\Omega|/2$ above, in which case the second term reduces to $T^{2}\Omega/12$.

\subsubsection{From chiral modes}

The contribution to the on-axis current from the chiral modes \eqref{eq:chiral-modes} due to rotation is given by
\begin{align}
   j_z^{\text{ax,ch}} 
   &= s\chi v_F\ell_z \int\frac{dk_z}{2\pi}\left(f_{\text{eq}}\left[E_{\text{rot}}^{\text{ch}}(n=0,k_z)\right]-f_{\text{eq}}\left[E_{\text{lab}}^{\text{ch}}(k_z)\right]\right)\left|\psi_{n=0,k_z,s}^\text{ch}(0,\phi,z)\right|^2 \\  
   &= \frac{s\chi v_F}{\pi R^2} \int\frac{dk_z}{2\pi}\left(f_{\text{eq}}\left[E_{\text{lab}}^{\text{ch}}(k_z)-\frac{s\Omega}{2}\right]-f_{\text{eq}}\left[E_{\text{lab}}^{\text{ch}}(k_z)\right]\right) \\
   &= \frac{\chi v_F}{(2\pi R)^2}\Omega
\end{align}
Above, we subtracted the current contribution in the absence of rotation to determine the rotation induced current. The restriction to $n=0$ comes from the fact that only such a mode is non-vanishing at $\rho=0$ according to \eqref{eq:chiral-modes}. $j_z^{\text{ax,ch}}$ clearly vanishes as $R\to\infty$, so the axial current in the thermodynamic limit is purely from the bulk wave-like states and matches the well-known semiclassical result.

\subsection{Cross-section average}\label{subsec:cross-section average}

Unlike the on-axis current, the cross-section average does not appear to have been analyzed previously and plays a central role in the interpretation developed in Sec. III.

\subsubsection{From bulk states}

The spatially averaged current from the wave-like bulk states is given by 
\begin{align}
{j_{z}^\text{avg,bulk}} & =\frac{1}{V}\int d^{3}rj_z^\text{bulk}(\rho)\\
& =\frac{1}{V}\sum_{\lambda,n,k_{\perp},k_{z}}f_{\text{eq}}\left(E_{\text{rot}}^{\lambda}\right)\left\langle \psi_{\lambda,n,k_{\perp},k_{z}}\left|\chi{v_F}\sigma_{z}\right|\psi_{\lambda,n,k_{\perp},k_{z}}\right\rangle \\
& =\frac{\chi}{V}{v_F}\int d^{3}r\sum_{\lambda,n,k_{\perp},k_{z}}f_{\text{eq}}\left[E_{\text{rot}}^{\lambda}(n,k_{\perp},k_z)\right]\left(\left|a_{\lambda}(k_{\perp},k_{z})\right|^{2}J_{n}^{2}(k_\perp\rho)-\left|b_{\lambda}(k_{\perp},k_{z})\right|^{2}J_{n+1}^{2}(k_\perp\rho)\right)\label{eq:jz-bulk-local}
\end{align}
For large enough $R$, $J_n(k_\perp\rho)$ and $J_{n+1}(k_\perp\rho)$ must have identical normalizations as described in Sec. \ref{subsubsec:bulk}. As a result, $j_z^\text{avg}$ simplifies to
\begin{align}
j_z^\text{avg,bulk} & =\chi\frac{1}{V}v_F\sum_{\lambda,n,k_{\perp},k_{z}}f_{\text{eq}}\left[E_{\text{lab}}^{\lambda}(n,k_{\perp},k_{z})-\left(n+\frac{1}{2}\right)\Omega\right]\left(\left|a_{\lambda}(k_{\perp},k_{z})\right|^{2}-\left|b_{\lambda}(k_{\perp},k_{z})\right|^{2}\right)\\
 & =\chi\frac{1}{V}v_F\sum_{\lambda,n,k_{\perp},k_{z}}f_{\text{eq}}\left[\lambda{v_F}\sqrt{k_{z}^{2}+k_{\perp}^{2}}-\mu-\left(n+\frac{1}{2}\right)\Omega\right]\frac{\lambda k_{z}}{\sqrt{k_{z}^{2}+k_{\perp}^{2}}}\\
 & =0,
\end{align}
since the integrand is odd in $k_{z}$. Since the average current
density vanishes, the current density calculated in Sec. \ref{subsec:rho0 current} from the bulk states is purely a magnetization current density.

To see this explicitly, we can construct the local magnetization density that produces $j_z^\text{bulk}(\rho) = (\boldsymbol{\nabla}\times\mathbf{M}(\mathbf{r}))_z$
with $\mathbf{M}(\rho,\phi,z)=M_\phi(\rho)\hat{\boldsymbol{\phi}}$ independent of $\phi,z$ and
\begin{align}
    M_\phi(\rho) &=\frac{1}{\rho}\intop_0^\rho \rho'd\rho' j_z(\rho') \nonumber\\
    &=\frac{\chi v_F\rho}{4}\sum_{\lambda,n,k_{\perp},k_{z}}f_{\text{eq}}\left[E_{\text{rot}}^{\lambda}(n,k_{\perp},k_z)\right]\left(J_n^2(k_\perp\rho)-J_{n-1}(k_\perp\rho)J_{n+1}(k_\perp\rho)-J_{n+1}^2(k_\perp\rho)+J_{n}(k_\perp\rho)J_{n+2}(k_\perp\rho)\right)\nonumber\\
    &=\frac{\chi v_F}{2}\sum_{\lambda,n,k_{\perp},k_{z}}\frac{1}{k_\perp}f_{\text{eq}}\left[E_{\text{rot}}^{\lambda}(n,k_{\perp},k_z)\right]J_n(k_\perp\rho)J_{n+1}(k_\perp\rho)\nonumber\\
\end{align}
after dropping terms that are odd in $k_z$. Using the asymptotic forms of the Bessel functions for small and large arguments, $M_\phi(\rho)\sim \rho^{2n+1}$ for small $k_\perp\rho$ and $M_\phi(k_\perp\rho)\sim1/\rho$ for large $k_\perp\rho$. In other words, $M_\phi(\rho)$ vanishes at both small and large distances and is a well-behaved function, so its curl averaged over a cross-section vanishes.

\subsubsection{From chiral modes}

To calculate the cross-section averaged current density due to rotation from the chiral modes, we must subtract the current density in the absence of rotation. Thus,

\begin{align}
    j_z^{\text{avg,ch}} 
    &= \frac{s\chi v_F}{V}\int d^3 r \sum_{n\geq0,k_z}\left(f_{\text{eq}}\left[E_{\text{rot}}^{\text{ch}}(n,k_z)\right]-f_{\text{eq}}\left[E_{\text{lab}}^{\text{ch}}(k_z)\right]\right)\left|\psi_{n,k_z,s}^\text{ch}(\rho,\phi,z)\right|^2 \\
    &= \frac{s\chi v_F}{\pi R^2} \sum_{n\geq0}\int\frac{dk_z}{2\pi}\left(f_{\text{eq}}\left[E_{\text{rot}}^{\text{ch}}(n,k_z)\right]-f_{\text{eq}}\left[E_{\text{lab}}^{\text{ch}}(k_z)\right]\right) \\
    &= \frac{s\chi v_F}{\pi R^2} \sum_{n\geq0}\int\frac{dk_z}{2\pi}
    \left(f_{\text{eq}}\left[E_{\text{lab}}^{\text{ch}}(k_z)-s\left(n+\frac{1}{2}\right)\Omega\right]-f_{\text{eq}}\left[E_{\text{lab}}^{\text{ch}}(k_z)\right]\right) \\
    &= \frac{\chi \Omega}{2\pi^2 R^2}\sum_{n\geq0}\left(n+\frac{1}{2}\right)\\
    &= \frac{\chi \Omega N^2}{(2\pi R)^2}
\end{align}
where $n=0,1,2,\dots,N-1$.

\begin{figure}
\includegraphics[scale=0.6]{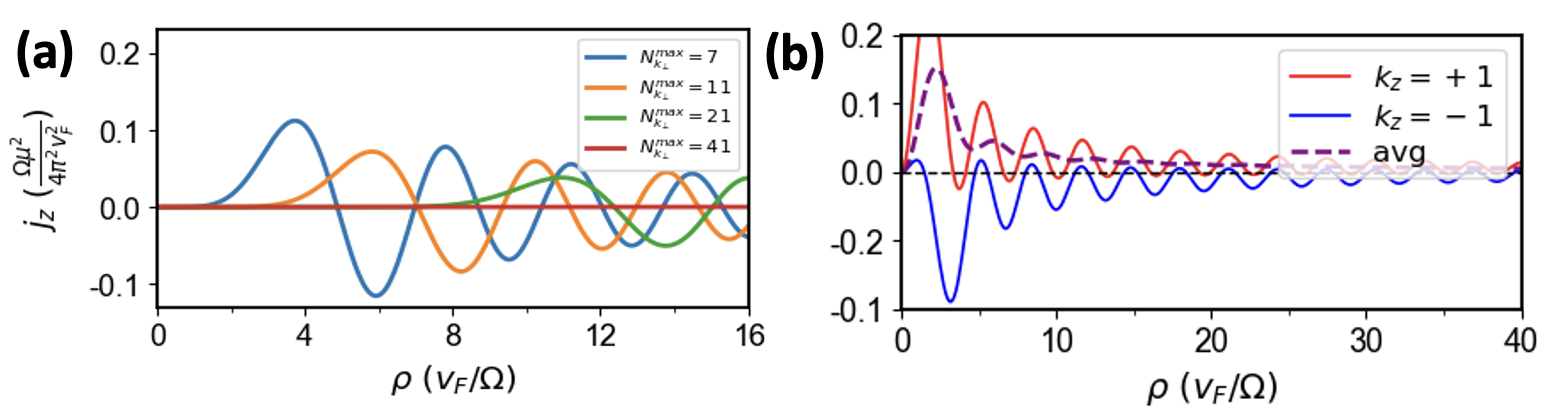}
\caption{\label{fig:epsart}(a) $j_z(\rho)$ from the bulk states for various cutoff choices $N_\text{max}$ on $n$. The contribution tends to locally vanish as $N_\text{max}$ grows due to cancellation between various oscillating functions. (b) The global cancellation (dashed) due to oddness of current density with respect to $k_z$, exemplified by $k_z={\pm}1$ modes. The dashed line is defined by $\int_0^{\rho}(j_{k_z=1}+j_{k_z=-1})$. \label{f3}}
\end{figure}

\subsection{On the surface}

\subsubsection{From bulk states}

The current density on the surface from the bulk wave-like states is 
\begin{align}
j_{z}^{\text{bulk}}(R) & =\chi{v_F}\sum_{\lambda,n}\intop_{\mathbf{k}}f_{\text{eq}}\left[E_{\text{rot}}^{\lambda}(n,k_{\perp},k_z)\right]\left(\left|a_{\lambda}(k_{\perp},k_{z})\right|^{2}J_{n}^{2}(k_{\perp}R)-\left|b_{\lambda}(k_{\perp},k_{z})\right|^{2}J_{n+1}^{2}(k_{\perp}R)\right)\\
 & =\frac{\chi}{2}v_F\sum_{\lambda,n}\intop_{\mathbf{k}}f_{\text{eq}}\left[E_{\text{rot}}^{\lambda}(n,k_{\perp},k_z)\right]\left[\left(1+\frac{\lambda k_{z}}{\sqrt{k_{z}^{2}+k_{\perp}^{2}}}\right)J_{n}^{2}(R)-\left(1-\frac{\lambda k_{z}}{\sqrt{k_{z}^{2}+k_{\perp}^{2}}}\right)J_{n+1}^{2}(R)\right]\\
 & =\frac{\chi}{2}v_F\sum_{\lambda,n}\intop_{\mathbf{k}}f_{\text{eq}}\left[E_{\text{rot}}^{\lambda}(n,k_{\perp},k_z)\right]\left(J_{n}^{2}(k_{\perp}R)-J_{n+1}^{2}(k_{\perp}R)\right)\\
 & \approx\chi{v_F}\sum_{\lambda,n}\intop_{\mathbf{k}}f_{\text{eq}}\left[E_{\text{rot}}^{\lambda}(n,k_{\perp},k_z)\right]\frac{(-1)^n\sin(2k_{\perp}R)}{\pi k_{\perp}R}
\end{align}
for $k_\perp R$ which vanishes as $R\to\infty$. Recall that for polarized boundary conditions \eqref{eq:polarized}, there are no surface states.

\subsubsection{From chiral modes}

To calculate the surface current density from the chiral states due to rotation, we must regularize by subtracting the current in the absence of rotation as usual. Thus,
\begin{align}
    j_z^{\text{ch}}(R) &= s\chi v_F\ell_z \sum_{n\geq0}\int\frac{dk_z}{2\pi}\left(f_{\text{eq}}\left[E_{\text{rot}}^{\text{ch}}(n,k_z)\right]-f_{\text{eq}}\left[E_{\text{lab}}^{\text{ch}}(k_z)\right]\right)\left|\psi_{n,k_z,s}^\text{ch}(R,\phi,z)\right|^2 \\
    &=
    \frac{s\chi v_F}{\pi R^2} \sum_{n\geq0}\int\frac{dk_z}{2\pi}
    \left(f_{\text{eq}}\left[E_{\text{rot}}^{\text{ch}}(n,k_z)\right]-f_{\text{eq}}\left[E_{\text{lab}}^{\text{ch}}(k_z)\right]\right)(n+1) \\
    &= \frac{s\chi v_F}{\pi R^2} \sum_{n\geq0}\int\frac{dk_z}{2\pi}
    \left(f_{\text{eq}}\left[E_{\text{lab}}^{\text{ch}}(k_z)-s\left(n+\frac{1}{2}\right)\Omega\right]-f_{\text{eq}}\left[E_{\text{lab}}^{\text{ch}}(k_z)\right]\right)(n+1) \\
    &= \frac{\chi \Omega}{2\pi^2 R^2}\sum_{n\geq0}\left(n+\frac{1}{2}\right)(n+1) \\
    &= \frac{\chi \Omega}{(2\pi R)^2}\frac{N(N+1)(4N-1)}{12}
\end{align}

\section{Topology of the chiral modes}
\label{sec:chiral-topology}

Section IV presented the physical construction of the boundary chiral modes and their associated pumping phenomenon. In this appendix we provide a self-contained mathematical treatment, including the uniqueness proof for the canonical Weyl node with arbitrary cross-sectional geometry and the extension to arbitrary spin $S$, Weyl charge $q$, where they constitute an exactly solvable family of solutions but may not be unique. Under rotation, all these chiral modes pump charge by an amount that ties together quantum geometry in momentum, real and spin space.

\subsection{Canonical Weyl case: exact chiral modes and their uniqueness}
\label{subsec:canonical-chiral}

For completeness we repeat the canonical derivation in full before generalizing it. We begin with the ordinary Weyl fermion of spin $S=\frac12$ and unit
Weyl charge $q=1$, for which the structure of the chiral modes can be
derived completely and uniquely. This is the physically most relevant
case and also the one in which the fully spin-polarized chiral family
is fixed uniquely by the boundary condition.

Consider the Weyl Hamiltonian \eqref{eq:Hlab} in the laboratory frame. We look for eigenstates with the chiral dispersion
\begin{equation}
E^{\rm ch}=s\chi v_F k_z-\mu .
\end{equation}
where $s=\pm1$. Substituting this into the Schr\"odinger equation and writing the spinor as
\begin{equation}
\psi(\mathbf r)=
\begin{pmatrix}
\psi_\uparrow(\mathbf r)\\
\psi_\downarrow(\mathbf r)
\end{pmatrix},
\end{equation}
gives a solution for $s=1$ that is consistent with the boundary being spin-up:
\begin{equation}
(k_x-i k_y)\psi_\uparrow=0,
\qquad
\psi_\downarrow=0 .
\label{eq:canonical-chiral-eqs}
\end{equation}
Thus, the chiral mode is fully spin polarized: the down-spin component
vanishes identically, while the up-spin component obeys a first-order
holomorphicity condition.

Introducing
\(
\zeta=x+iy
\)
and using
\(
k_x-i k_y=-i(\partial_x-i\partial_y)=-2i\,\partial_{\bar\zeta},
\)
Eq.~\eqref{eq:canonical-chiral-eqs} becomes
\begin{equation}
\partial_{\bar\zeta}\psi_\uparrow=0 .
\end{equation}
Hence $\psi_\uparrow$ is holomorphic in the transverse coordinates. On
a simply connected cross-section $D$, every regular holomorphic
function generates a chiral solution,
\begin{equation}
\psi(\mathbf r)
=
\frac{e^{ik_z z}}{\sqrt{\ell_z}}
\begin{pmatrix}
f(\zeta)\\
0
\end{pmatrix},
\qquad
\partial_{\bar\zeta}f=0 .
\label{eq:canonical-general-solution}
\end{equation}
For the circular cross-section, a convenient basis is provided by the
monomials
\begin{equation}
f(\zeta)=\zeta^n=(x+iy)^n,
\qquad n\ge 0,
\end{equation}
which are precisely the modes quoted in the main text. Each
nonnegative integer $n$ produces one chiral branch. In polar
coordinates,
\begin{equation}
\zeta^n=\rho^n e^{in\phi},
\end{equation}
so these states carry spatial angular momentum $l=n$. They can be
understood as the $k_\perp\to0$ limits of the bulk modes built from
Bessel functions $J_n$; in particular, the uniform chiral mode
corresponds to $n=0$ and follows from $J_n(0)=\delta_{n0}$:
\begin{equation}
\psi(\mathbf r)
=
\frac{e^{ik_z z}}{\sqrt{A\ell_z}}
\begin{pmatrix}
1\\
0
\end{pmatrix}.
\end{equation}

We now show that this fully spin-up chiral family is in fact unique.
Indeed, one may view the two spin components as satisfying decoupled
first-order equations, with the down-spin component obeying a
holomorphic or anti-holomorphic equation together with the boundary
condition $\psi_\downarrow=0$ on the boundary. By analyticity, a function of this type that vanishes on the entire boundary must vanish identically in
the interior. Therefore,
\begin{equation}
\psi_\downarrow(\mathbf r)\equiv 0
\qquad \text{throughout } D .
\end{equation}
It follows that every chiral mode consistent with the fully
spin-polarized boundary condition must be of the form
\eqref{eq:canonical-general-solution}. In this canonical case
$S=\frac12$, $q=1$, the exactly solvable family of chiral modes is
therefore unique.

This clean uniqueness is special to the ordinary Weyl case. As we show
next, exact fully spin-polarized chiral families continue to exist for
more general spin $S$, Weyl charge $q$, and cross-section shape, but
their uniqueness is no longer guaranteed.

\subsection{Beyond $S=1/2$, unit Weyl charge and circular symmetry}
\label{subsec:robustness}

Having established the canonical Weyl case, we now extend the construction beyond spin-$1/2$, unit Weyl charge, and circular symmetry. We first show that the structure responsible for the chiral branches does not rely on the special assumptions of spin-$1/2$, unit Weyl charge, or circular symmetry of the boundary.  The argument applies to a general cross–section and to Weyl nodes with arbitrary charge $q$ and spin $S$.

Consider a generalized Weyl Hamiltonian
\begin{equation}
H_{\rm lab}
=
2\chi\!\left(
v k_z S_z
+
u k_-^{\,q} S_+
+
u k_+^{\,q} S_-
\right)
-\mu ,
\end{equation}
where $k_\pm = k_x \pm i k_y$ and $S_\pm$ are spin ladder operators.
This form captures Weyl nodes of arbitrary topological charge $q$
as well as higher–spin generalizations protected by crystalline
symmetry. We impose a spin–polarized boundary condition that fixes the spin
projection on the boundary $\partial D$ of the transverse region $D$,
\begin{equation}
S_z\psi|_{\partial D}=s\,\psi|_{\partial D}.
\end{equation}
In particular, when $s=\pm S$ the boundary selects the extremal spin
state.  Importantly, this boundary condition constrains only the spin
degree of freedom and places no restriction on the spatial dependence
of the wavefunction along the boundary.

We now look for eigenstates whose dispersion is linear in $k_z$:
\begin{equation}
E_{\rm ch}=2\chi Sv k_z-\mu .
\end{equation}
Substituting this form into the Schr\"odinger equation shows that the
transverse wavefunction must satisfy
\begin{equation}
\left(
k_-^{\,q} S_+
+
k_+^{\,q} S_-
\right)
\psi(\mathbf r_\perp)=0 .
\end{equation}
If the boundary selects the extremal spin state $|S\rangle$, the
ladder operator annihilates it,
\begin{equation}
S_+|S\rangle=0 .
\end{equation}
Consequently there exists solutions of the form
\begin{equation}
\psi(\mathbf r_\perp)=w(\zeta,\bar\zeta)\,|S\rangle ,
\qquad
\zeta=x+iy .
\end{equation}
where the spatial and spin parts have factorized. The spatial part satisfies
\begin{equation}
k_-^{\,q}w(\zeta,\bar\zeta)
=
\left[-i(\partial_x-i\partial_y)\right]^q
w(\zeta,\bar\zeta)
=[-i(\partial_{\bar\zeta})]^q w(\zeta,\bar\zeta)=0
\end{equation}
The general normalizable solution is therefore a function of the form
\begin{align}
w(\zeta,\bar\zeta)
&=
\bar\zeta^m\zeta^n\qquad\text{for}\quad0\leq m < q,\quad\max(0,-N+m)\leq n<N+m,\quad m,n\in\mathbb{Z} \\
&=\rho^{m+n}e^{i(n-m)\phi}
\end{align}
upto normalization, which has spatial angular momentum $l=n-m$. Regularity requires $m+n\geq0$ while a spatial angular momentum cutoff $N$ implies $l=n-m\in(-N,N)$. Each such transverse profile produces a chiral eigenstate
\begin{equation}
\psi_{\rm ch}(\mathbf r)
=
\frac{e^{ik_z z}}{\sqrt{\ell_z}}
\,w(\zeta,\bar\zeta)\,|S\rangle ,
\end{equation}
with dispersion $E_{\rm ch}=2S\chi v k_z-\mu $.

Thus the existence of chiral branches follows directly from the
structure of the Hamiltonian and the spin–polarized boundary
condition, without requiring circular symmetry of the boundary or
restricting to $S=1/2$ or $q=1$.  The simplest member of this family for given $S,q$ corresponds to the spatially uniform solution $w=\mathrm{const}$, while more general functions generate additional chiral modes distinguished by their transverse structure.

\subsection{Quantized charge pumping}
\label{subsec:pumping}

We showed in App. \ref{subsec:cross-section average} that the chiral modes in a unit Weyl fermion with $q=1$ and spin $S=1/2$ produce a spatially averaged current
\begin{equation}
    j_z^{\text{avg,ch}} 
    = \frac{\chi \Omega N^2}{4\pi A}
\end{equation}
where we have replaced $\pi R^2$ by the cross-sectional area $A$. The arguments of the previous subsection imply that this result does not depend on the shape of the cross-section but only depends on its area $A$. The total current is
\begin{equation}
    I^\text{ch}=Aj_z^{\text{avg,ch}}=\frac{\chi \Omega N^2}{4\pi}
\end{equation}
Furthermore, integrating over time gives the charge pumped by rotation:
\begin{equation}
\Delta Q^{\text{ch}}=\chi\frac{\Delta\theta}{4\pi}N^2
\end{equation}
where $N\geq1$ is the number of chiral modes. This is a remarkable result, stating that a minimum of one particle is pumped along $z$ for every two full rotations. Although it is a quantum phenomenon, it is independent of $\hbar$. The result is solely dependent on the periodicity of spatial rotation and intrinsic nature of the chiral fermions. In general, however, the charge pumped depends on the UV cutoff on angular momentum $N$.

The results in App. \ref{subsec:robustness} allow us to immediately compute the contribution to the charge pumped by the exactly solvable modes for general $S$ and $q$. Explicitly,
\begin{align}
    I^{\text{ch}}
    &= \frac{\chi \Omega S}{\pi}\sum_{m=0}^{q-1}\sum_{n=\max(0,-N+m)}^{N+m}\left(n+\frac{1}{2}\right)\\
    &= \frac{\chi \Omega q S}{12\pi}\left[6 N^2+6N(q-1)+(2q-1)(q-1)\right]\\
    &\approx \frac{\chi \Omega q S N^2}{2\pi}
\end{align}
for $N\gg q$. Thus, the pump simply gets multiplied by factors $q$ and $S$ and pumps charge
\begin{equation}
\Delta Q^{\text{ch}}=\chi q S\frac{\Delta\theta}{2\pi}N^2
\end{equation}

\end{widetext}

\bibliography{library}
\bibliographystyle{apsrev4-2}
\end{document}